\documentclass[11pt]{article}
\usepackage{graphicx}

\begin{document}

\title{A Mirror Based Event Cloaking Device}

\author{Miguel A. Lerma}

\maketitle

\begin{abstract}
We propose a way of implementing an event cloaking device without the
use of metamaterials.  Rather than slowing down and speeding up light,
we manipulate an obscurity gap by diverting the light through paths of
appropriate length with an arrangement of switchable transreflective
mirrors.

\end{abstract}



\section{Introduction}

A spacetime cloak, or event cloak, is a means of manipulating
electromagnetic radiation in space and time in such a way that a
certain collection of happenings, or events, is concealed from distant
observers. Conceptually, a safecracker can enter a scene, steal the
cash and exit, whilst a surveillance camera records the safe door
locked and undisturbed all the time. 

An event cloak design using metamaterials was first proposed
theoretically by a team of researchers from Imperial College London
(UK) in 2010, and published in the Journal of Optics
\cite{maccal:2011}. Their design works by using a medium in which
different parts of the light illuminating a certain region can be
either slowed down or speed up. A leading portion of the light is
speeded up so that it arrives before the events occur, whilst a
trailing part is slowed down and arrives too late. After their
occurrence, the light is reformed by slowing down the leading part and
speeding up the trailing part. The distant observer therefore only
sees a continuous illumination, whilst the events that occurred during
the dark period of the cloak's operation remain undetected. An
experimental demonstration of the basic concept using nonlinear
optical technology has been presented in a preprint on the Cornell
physics arXiv \cite{fridman:2011}.

Here we describe a similar event cloak device without metamaterials,
using only a system of mirrors that create a temporary gap of
obscurity, and close that gap afterward.

\section{Description of the mirror-based event cloaking device.}

\begin{figure}[htb]
\begin{center}
\ \includegraphics[height=1.5in]{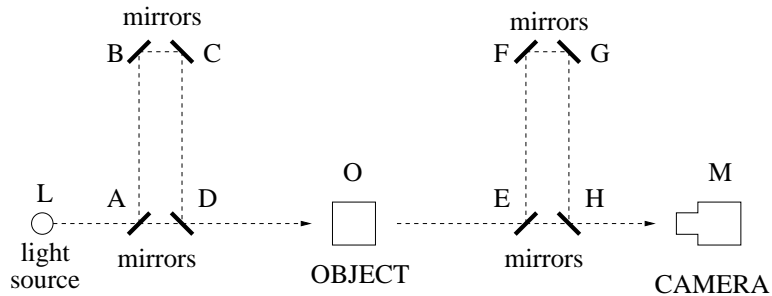}
\caption{Basic design of mirror-based event cloak.}\label{f:cloakbasic}
\end{center}
\end{figure}

Figure~\ref{f:cloakbasic} shows the basic arrangement for a
mirror-based event cloak device.  It consists of a light source L, and
a system of mirrors A, B, C, D, E, F, G, H, of which A, D, E and H are
switchable between several possible states: fully transparent (letting
light go through), fully reflective (working as an ordinary mirror),
and adjustable degrees of half-reflection.  Electrically switchable
transreflective mirrors are currently available, so the device
described here is within the scope of current technology.

The arrangement of the mirrors may change, for instance it is possible
to make the paths ABCD and EFGH longer by inserting extra mirrors (and
so obtain a larger time gap for the event cloaking effect), but the
times taken by the light to go through each of those paths must be
identical.

The light produced by the light source L illuminates the object O
after following one of the two paths LADO, or LABCDO, depending on
whether mirrors A and D are set in the transparent or the reflective
state. Then, the light leaving the object will reach the camera M also
after following one of two paths, OEHM, or OEFGHM, depending on the
transparent or reflective state of mirrors E and H.

\section{Performing event cloaking.}

In order to accomplish the event cloak effect, switching of the four
mirrors A, D, E, and H must be carefully timed, so to create a
temporary gap of obscurity in the light arriving to the object, to be
precisely closed in the light leaving the object and arriving to the
camera.

\begin{figure}[htb]
\begin{center}
\ \includegraphics[height=1.5in]{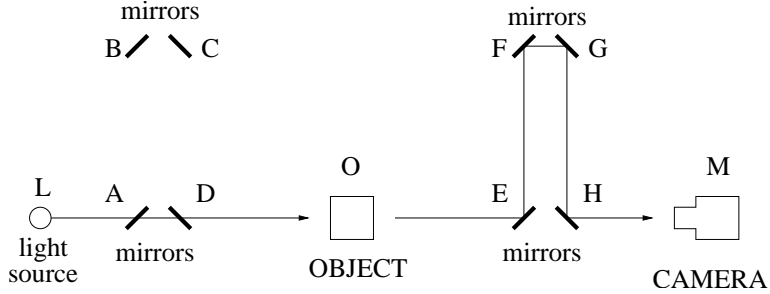}
\caption{Initial setting: A and B are transparent, E and H are reflective.}\label{f:cloakset}
\end{center}
\end{figure}

The system is initially set as shown in figure~\ref{f:cloakset},
with mirrors A and D in their transparent state, and mirrors E and H
in their reflective state.

\begin{figure}[htb]
\begin{center}
\ \includegraphics[height=1.5in]{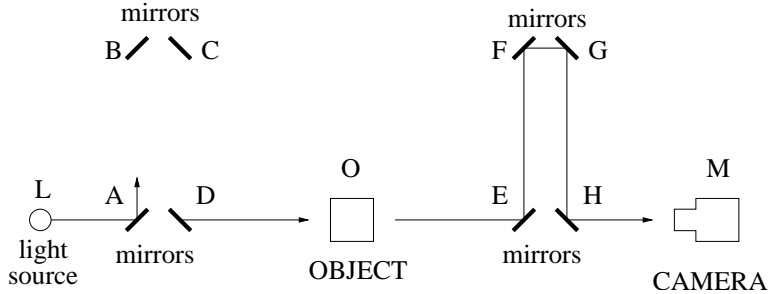}
\caption{Starting the obscurity gap: A switches to a reflective state, a little later D switches to reflective too.}\label{f:cloakgapstart}
\end{center}
\end{figure}

The event cloaking effect starts as shown in
figure~\ref{f:cloakgapstart}, with mirror A switching to its
reflective state, and after the light between A and D has gone through
mirror D, this mirror switches to a reflective state too.  This create
an obscurity gap of duration equal to the time taken by the light to
go through the path ABCD minus the time taken to go directly from A to
D---in the figure AD=BC and AB=CD, so the duration of the gap would be
twice the time taken by the light to go from A to B.

\begin{figure}[htb]
\begin{center}
\ \includegraphics[height=1.5in]{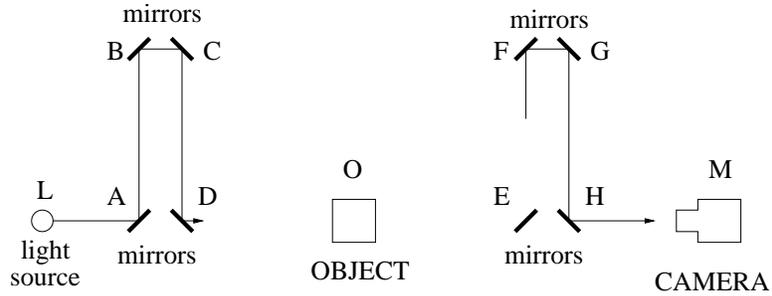}
\caption{The object in the middle of the obscurity gap.}\label{f:cloakobscure}
\end{center}
\end{figure}

Figure~\ref{f:cloakobscure} shows the object in total obscurity.
Anything that happens at O during the time duration of the obscurity
gap will be invisible for the camera.

\begin{figure}[htb]
\begin{center}
\ \includegraphics[height=1.5in]{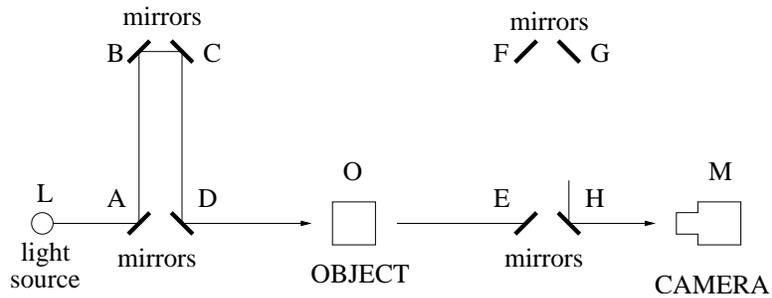}
\caption{End of the obscurity gap. By now E has switched to transparent.}\label{f:cloakgapend}
\end{center}
\end{figure}

Figure~\ref{f:cloakgapend} shows the end of the obscurity gap.  The
object is being illuminated again and the closing of the gap starts by
switching mirrors E and (a little later) H to transparent state.

\begin{figure}[htb]
\begin{center}
\ \includegraphics[height=1.5in]{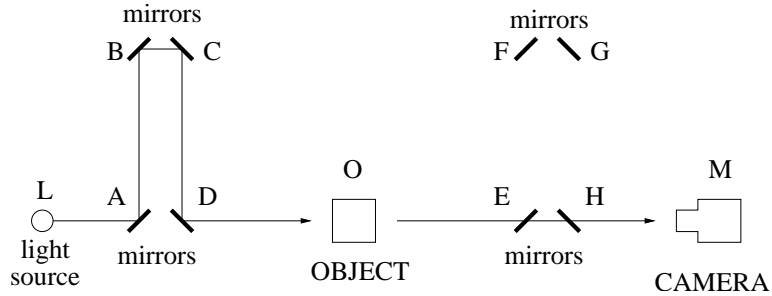}
\caption{Obscurity gap closed, E and
H are both transparent, event cloak finished.}\label{f:cloakfinish}
\end{center}
\end{figure}

In figure~\ref{f:cloakfinish} the obscurity gap has been closed, E and
H are both transparent, event cloaking finished.

During all this time the camera has not registered any interruption in
the reception of the image of the object, although nothing happening
at O during the obscurity gap has been recorded by the camera.  The
only clue of the cloaking phenomenon would be a sudden jump in time in
the image received by the camera.  If for instance there is a clock at
O showing 12:00~pm at the moment in which the obscurity gap reaches
the object, and the gap lasts 5 minutes, then the image recorded by the
camera would register a sudden jump from 12:00~pm to 12:05~pm.

\section{Resetting the device.}

If we want to use the device again we need to reset it to its original
state shown in figure~\ref{f:cloakset}.  In order to do so we time
the transreflective mirrors A, D, E, and H to switch in the way
described below.

\begin{figure}[htb]
\begin{center}
\ \includegraphics[height=1.5in]{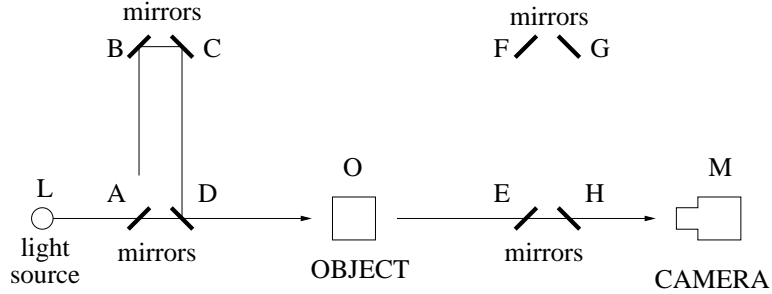}
\caption{Reset starts: A switches to transparent.}\label{f:cloakresetstart}
\end{center}
\end{figure}

First we switch mirror A to its transparent state, as shown in
figure~\ref{f:cloakresetstart}.

\begin{figure}[htb]
\begin{center}
\ \includegraphics[height=1.5in]{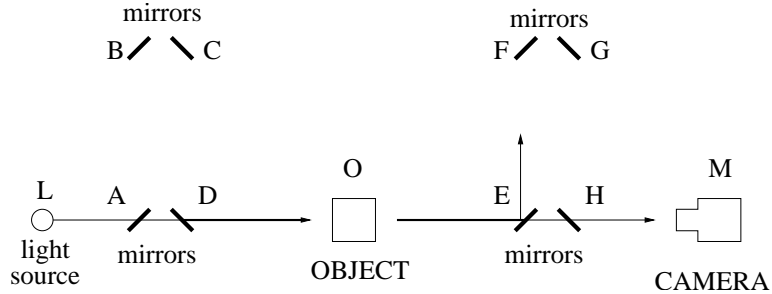}
\caption{Beams combined at D illuminate the object. Mirror E working as a splitter.}\label{f:cloakcombine}
\end{center}
\end{figure}

Then, for a time equal to the previous duration of the obscurity gap,
light going through the path AD will arrive at D at the same time as
light that took the path ABCD.  Combining the two beams into one may
have different effects depending on the kind of light used.  With
ordinary light we may obtain a light beam with roughly the sum of the
intensities of the incident beams, but other kinds of light (such as
laser) may cause interferences.  Here we leave open the precise way to
combine the incident beams and its consequences, and will assume for
now that they produce a (double intensity) combined beam leaving D
towards O, as shown in figure~\ref{f:cloakcombine}.

As soon as the light in the path ABCD has exited, mirror D can be
switched to a fully transparent state.  Also, during the time the
combined beam is arriving at E, this mirror must work as a splitter,
producing two beams, one going directly from E to H, and another one
following the path EFGH (figure~\ref{f:cloakcombine}).

\begin{figure}[htb]
\begin{center}
\ \includegraphics[height=1.5in]{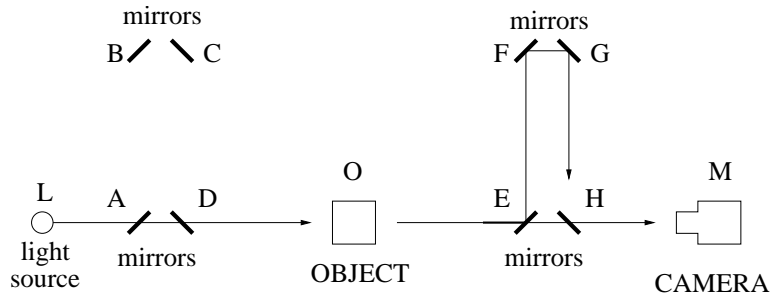}
\caption{Reset finishing. Mirror E switches to fully reflective, and a little later H does the same.}\label{f:cloakresetending}
\end{center}
\end{figure}

\begin{figure}[htb]
\begin{center}
\ \includegraphics[height=1.5in]{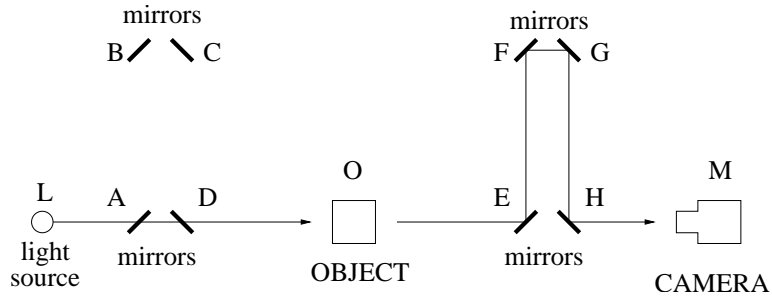}
\caption{Reset done.}\label{f:cloakresetdone}
\end{center}
\end{figure}

Figure~\ref{f:cloakresetdone} shows the end of the reset process.
Note that the light arriving at E has been split into two beams that
will arrive at M at different times, so the camera will witness the
object O going through the same period of time twice.  If there is for
instance a clock showing 12:30~pm at the moment the combined beam
created at D arrives at O, and that bean illuminates the object for 5
minutes (same as the obscurity gap before), then the camera will
record the clock going from 12:30~pm to 12:35~pm, then jumping back to
12:30~pm, and working normally from then on.

\section{Conclusions}

We have shown how to create an event cloak device without the use of
metamaterials, by a simple arrangement of switchable transreflective
mirrors. In the light arriving to an object an obscurity gap is
created by diverting the incoming light through a longer path, and
this gap is closed in the light path leaving the object by deviating
the light through a shorter path.

%


\end{document}